\documentclass[twocolumn,preprintnumbers,amsmath,amssymb]{revtex4}
\usepackage{latexsym,amssymb,amsthm,amsmath,epsfig}
\usepackage{hyperref}
\hypersetup{
    colorlinks=true,
    linkcolor=blue,
    filecolor=blue,      
    urlcolor=blue,
}

\begin{document}

\title{Quantitative analysis of $p$-wave three-body losses via cascade process}
\author{Muhammad Waseem$^{1}$}
\email{mwaseem328@gmail.com}
\author{Jun Yoshida$^{2,3}$}
\author{Taketo Saito$^{2,3}$}
\author{Takashi Mukaiyama$^{4}$}
\affiliation{%
$^{1}$\mbox{Karachi Institute of Power Engineering, Pakistan Institute of Engineering and Applied Sciences (PIEAS) } \\
$^{}$\mbox{Islamabad, Karachi, 75780, Pakistan}\\
$^{2}$\mbox{Department of Engineering Science, University of Electro-Communications, Tokyo 182-8585, Japan}\\
$^{3}$\mbox{Institute for Laser Science,University of Electro-Communications, Chofugaoka, Chofu, Tokyo 182-8585, Japan}\\
$^{4}$\mbox{Graduate School of Engineering Science, Osaka University, 1-3, Machikaneyama, Toyonaka, Osaka 560-8531 Japan}\\
}
\date{\today }

\begin{abstract}
We describe the three-body loss coefficient of identical fermions with $p$-wave interactions using a set of rate equations in which three-body recombination happens via an indirect process.
Our theoretical treatment explains experimental results just above the universal scaling law regime of weak interactions. 
Furthermore, we theoretically extend and experimentally verify the rate equation model for the case of atoms trapped in two dimensions.
Moreover, we find that the three-body loss coefficient in a two-dimensional trap is proportional to $A_{p}^{3}$ in the weakly interacting regime, where $A_{p}$ is the scattering area. Our results are useful in understanding three-body physics with $p$-wave interactions.
\end{abstract}

\maketitle

\section{introduction}
Many-body quantum systems such as ultracold atoms offer great control and versatility due to Feshbach resonances, which allow accurate control of interatomic interactions via external magnetic fields.
Such resonances occur when the kinetic energy of two colliding atoms in an open channel and the energy of bound-state atoms in the closed channel potential become degenerate.
Higher partial wave interactions across these resonances, such as $p$-wave Feshbach resonances, have specific features and signatures unlike their $s$-wave counterparts~\cite{vg, cheng, sukjin, victor, lev}, for example, anisotropic phases of $p$-wave condensates~\cite{cheng, lev} and the possibility of observing different quantum phase transitions across $p$-wave Feshbach resonances due to rich $p$-wave order parameters~\cite{vg, victor, lev}.
Some of these phase transitions across the $p$-wave Feshbach resonances for fermions are similar to the phases of superfluid $^3$He~\cite{lev11}.

The $p$-wave Feshbach resonances in ultracold fermionic gases have been observed experimentally~\cite{regal, zhang}. These $p$-wave Feshbach resonances have been widely used to explore inelastic collision losses~\cite{regal, zhang, ticknor, chevy, gun, fed, russian, waseem2, yoshida, waseem3,elk}, the creation of $p$-wave Feshbach molecules~\cite{zhang, gab, inada, waseem}, the binding energies of $p$-wave Feshbach molecules~\cite{gab, fuch}, the determination of scattering parameters~\cite{naka}, and $p$-wave contacts~\cite{ueda1, thy, yao, lio}. 
Following the advancement of $s$-wave few-body physics, an impressive amount of theoretical effort was devoted to the case of identical Fermions near a $p$-wave Feshbach resonance in three dimensions~\cite{esry1, suno, suno2, jona, jesper, jpd, bratpra} and in two dimensions~\cite{lev,nishida, nishida12, nishida13, nishida14, nikolaj, marish, jpd2d, zen1, zen2, maxim, sjj, hui, petrov2}.
On the experimental side, three-body loss coefficients $K_{3}$ as a function of interaction strength and temperature were measured in three-dimensional (3D) traps~\cite{regal, zhang, yoshida, waseem3}.
The experiments conducted in Ref.~\cite{yoshida} successfully verified the temperature threshold behavior of $K_{3} \propto T^{2}$ at the low-temperature ($T$) limit and the scattering volume scaling law of three-body loss coefficients, $K_{3} \propto V_{p}^{8/3}$, but only for relatively weak interactions.
Since the first experimental study of spin-polarized interacting Fermi gases in two dimensions using a $p$-wave Feshbach resonance~\cite{gun}, systematic studies of two-body spin relaxation losses~\cite{waseem2} and the creation of the $p$-wave Feshbach molecules in the $m_{l}=\pm 1$ angular momentum state have been performed~\cite{waseem}. A systematic understanding of three-body collision losses of identical fermions confined in a two-dimensional (2D) trap is also needed in experiments due to its importance in stability~\cite{fed, lev} and few-body physics~\cite{nishida,nikolaj}.

In this paper, we present a quantitative analysis to explain the three-body loss coefficient of identical fermions with $p$-wave interactions using a set of rate equations describing three-body recombination as an indirect process.
Our theoretical model explains the experimental results only above the universal scaling law regime of weak interactions and requires only a single parameter representing a collision rate coefficient between a free atom and quasiresonant molecules. For cross verification of the rate equation model, we extend it to the case of atoms trapped in two dimensions and again find good agreement with experimental data above the weak interaction regime. 
Furthermore, we found that the three-body loss coefficient in a 2D trap is proportional to $A_{p}^{3}$ in the weakly interacting regime, where $A_{p}$ is the scattering area.

This article is organized as follows. Section II describes our theoretical treatment for three-body recombination losses as an indirect process. 
In Sec. III, we compare our theoretical discussion with experimental results.
In Sec. IV, we present our conclusions and outlook regarding this work.

\section{Theoretical Model}
\emph{Three-dimensional theory:} 
We consider the narrow Feshbach resonance that occurs in a $p$-wave collision between the lowest-energy hyperfine state of $^6$Li atoms located at a magnetic field strength of $B_0=$~159.17(5)~G.
The two-body $p$-wave interactions are described by an effective-range expansion of the $p$-wave scattering phase shift $\delta_{p}(k)$ as~\cite{Taylor, naidon}
\begin{equation}
k^{3} \cot \delta_{p}(k)= -1/V_{p}-k_{\rm e} k^{2},
\label{range3d}
\end{equation}
where $k$ is a relative wave vector and $k_{\rm e}$ is the effective range defined as positive.
The scattering volume $V_{p}$ plays a role analogous to that of $s$-wave scattering length $a$ and varies as a function of the external magnetic field $B$ as $V_{p}=V_{\rm bg}[1-\Delta B/(B-B_{0})] \approx V_{\rm bg} \Delta B/(B-B_{0}) $.
Here, $V_{\rm bg}$ and $\Delta B$ are the background scattering volume and resonance width in the magnetic field, respectively~\cite{idz}.
We used the values $V_{\rm bg}\Delta B=-(2.8 \pm 0.3) \times 10^{6} a_{0}^{3}$~[${\rm G} {\rm m}^{3}$] and $k_{\rm e} = (0.055 \pm 0.005)a_{0}^{-1}$ from our previous measurement~\cite{naka}, where $a_{0}$ is the Bohr radius.
The energy of the molecular state relative to the threshold varies as $E_{\rm r}=\hbar^{2} k_{\rm r}^{2}/m$ where $m$ is the mass of a $^6$Li atom and $k_{\rm r}=1/\sqrt{\left|V_{p}\right| k_{\rm e}}$ is the momentum of the resonant bound state parametrized by the scattering volume $V_{p}$ and the effective range $k_{\rm e}$~\cite{russian}.
The relative kinetic energy $E=\hbar^{2} k_{\rm T}^{2}/m$ becomes equal to the energy of the molecular state relative to the threshold $E_{\rm r}$ at resonance, where $k_{\rm T}=\sqrt{3m k_{\rm B} T/(2\hbar^2)}$ is the thermal relative wave number.
Near the threshold, the finite lifetime of the molecular state due to coupling to the continuum is described by the energy-dependent resonance width $\Gamma_{\rm r}=(2 \sqrt{m} /k_{\rm e} \hbar) E^{3/2}$~\cite{victor, thy}.

In order to describe three-body recombination near the narrow $p$-wave Feshbach resonance, we treated the loss of trapped atoms as an indirect successive process.  
A $p$-wave Feshbach resonance can be considered as a bound quasimolecular state weakly coupled to a continuum above the resonance~\cite{victor}.
Therefore, the rate coefficient for the generation of quasimolecules $K_{M}$ via two-body collision at temperature $T$ above the narrow $p$-wave Feshbach resonance is~\cite{lee}
\begin{equation}
K_{M}= (\Gamma_{\rm r}/ \hbar ) (6 \pi  /k_{\rm T}^{2})^{3/2} (2l+1) e^{-\left(k_{\rm r}/k_{\rm T}\right)^{2}},
\label{km}
\end{equation}
where $l=1$.
Then, the rate coefficient of vibrational quenching of a molecule, $K_{AD}$, occurs through collision of the molecule with a third atom. The rate equations describing the time evolution of atomic number density $n$ and the density of quasiresonant molecules $n_{D}$ in the resonance state are 
\begin{subequations}
\begin{align}
\frac{dn}{dt} & = 2 (\Gamma_{\rm r}/\hbar)n_{D}-2 K_{M} n^{2}-K_{AD} n n_{D},\\
\frac{dn_{D}}{dt} & = -(\Gamma_{\rm r}/\hbar)n_{D}+K_{M} n^{2}-K_{AD} n n_{D}.
\label{rateb}
\end{align}
\end{subequations}
Here, instead of having a $- K_3 n^3$ term, with $K_3$ being the three-body loss coefficient in the rate equation for atomic density as a three-body loss term,
the three-body loss is taken into account as a successive process of dimer creation followed by an atom-dimer collision.
Assuming a steady-state condition of molecular density ($dn_{D}/dt=0$), one obtains
\begin{equation}
n_{D}=\frac{K_{M}}{\Gamma_{\rm r}/\hbar+n K_{AD}} n^{2}.
\end{equation}
Under the further condition of fast molecular creation, $(\Gamma_{\rm r}/ \hbar) \gg n K_{AD}$, we obtain a steady-state rate equation of the atomic density $n$ 
in the following form:
\begin{equation}
\frac{dn}{dt}=-\frac{3 K_{AD} K_{M}}{(\Gamma_{\rm r} / \hbar)} n^{3}=-K_{3} n^{3}.
\label{kn}
\end{equation}
Substituting Eq.~(\ref{km}) into Eq.~(\ref{kn}), we write the three-body loss coefficient $K_{3}$ in the following form:
\begin{equation}
K_{3} \approx 9 K_{AD} (6 \pi / k_{\rm T}^{2})^{3/2} e^{-\left(k_{\rm r}/k_{\rm T}\right)^{2}}.
\label{k3fit2}
\end{equation}
We assume that molecular creation is sufficiently fast that the three-body loss coefficient is directly related to the atom-dimer collision coefficient  $K_{AD}$.

\emph{Quasi-2D theory:}
Next, we apply a similar theoretical treatment to the case of atoms trapped in a quasi-two-dimensional geometry. In quasi-2D geometry, atoms are confined in the axial ($z$) direction with frequency $\omega_{z}$, and the interatomic separations in the other two directions ($x$ and $y$) greatly exceed the axial extension of the atomic wave function given by the harmonic oscillator length $l_{z}=\sqrt{\hbar / (m \omega_{z})}$.
In this case, the two-body $p$-wave interactions are described by a two-dimensional effective range expansion of the scattering phase shift $\delta_{p}(q)$ as~\cite{cai, prico, russian}
\begin{equation}
q^{2} \cot \delta_{p}(q)=-1/A_{p}-B_{p} q^{2},
\end{equation}
where $q$ is a relative wave vector defined for the 2D case.
The scattering area $A_{p}$, defined as $A_{p}=(3 \sqrt{2 \pi} l_{z}^{2}/4 ) / \left[(l_{z}^{3}/V_{p}) +(k_{\rm e} l_{z}/2)-0.065553 \right]$, was used as a controllable length scale parameter to tune the interaction via external magnetic field. $A_{p}$ depends upon both the harmonic oscillator length $l_{z}$ and the 3D scattering parameters~\cite{russian}.
The positive dimensionless effective range in two dimensions is $B_{p}=(4/3 \sqrt{2 \pi}) \left[k_{\rm e} l_{z}-0.14641 \right]-2 \ln (l_{z} q)/ \pi$, which also depends on the 3D effective-range parameter and the harmonic oscillator length~\cite{russian}.

For the 2D case, the rate for the generation of quasimolecules $Q_{M}$ via two-body collisions at temperature $T$ is written as
\begin{equation}
Q_{M}= (\gamma_{\rm r}/ \hbar) (4 \pi / q_{T}^{2}) (2l) ~ e^{-\left(q_{\rm r}/q_{\rm T}\right)^{2}}.
\label{Qm}
\end{equation}
Here, $q_{\rm T}=\sqrt{m k_{\rm B} T/\hbar^2}$ is the thermal momentum in quasi-2D geometry and $q_{\rm r}=1/\sqrt{\left|A_{p}\right| B_{p}}$ is the momentum of the resonant bound state parametrized by the scattering area $A_{p}$ and the effective range $B_{p}$~\cite{russian}. The two-dimensional resonance energy width is $\gamma_{\rm r}=2 E / B_{p}$, which can be obtained from the pole of the $p$-wave scattering amplitude $f_{p}(q)=-4/\left[\cot \delta_{p}(q)-i\right]$ in two dimensions~\cite{marish}.
Similar to the 3D case, we assume a steady-state condition for the molecular density and a condition of $(\gamma_{\rm r}/ \hbar) \gg Q_{AD} n_{s}$. 
Here, we denote the rate constant of vibrational quenching of molecules due to atom-dimer collisions with $Q_{AD}$ and atomic density with $n_{s}$. 
Then, the three-body loss rate coefficient can be obtained in the form
\begin{equation}
Q_{3} \approx \frac{24 \pi Q_{AD}}{q_{\rm T}^{2}}   ~ e^{-\left(q_{\rm r}/q_{\rm T}\right)^{2}},
\label{Q3fit}
\end{equation}
which again describes the three-body loss rate coefficient as a function of magnetic field detuning and the atomic temperature in terms of the atom-dimer rate coefficient $Q_{AD}$.

\section{Experiments}
\emph{Three-dimensional case:}
We first trapped $^{6}$Li atoms in the hyperfine ground state of $\left\vert F,m_{F}\right\rangle= \left\vert 1/2,1/2\right\rangle (\equiv \left\vert 1\right\rangle )$ and $\left\vert F,m_{F}\right\rangle =  \left\vert 1/2,-1/2 \right\rangle (\equiv \left\vert 2\right\rangle )$ using a single-beam optical dipole trap as described in detail elsewhere~\cite{naka}. In brief, we performed evaporative cooling at 300~G magnetic field where the elastic collision cross section is at its maximum after capturing the atoms with a single-beam optical dipole trap. To prepare a single-component Fermi gas, we eliminated the atoms in the $\left\vert 2\right\rangle$ state by shining the resonant light for a duration of 50~$\mu$s.
Then, we ramped the magnetic field close to the $p$-wave Feshbach resonance for the $\left\vert 1\right\rangle - \left\vert 1\right\rangle$ state located at 159.17~G to investigate the loss features.
Here, we reduced the fluctuation in the magnetic field to 8~mG by stabilizing the current in the Feshbach coils using a bypass circuit added in parallel to the coils~\cite{naka}. 
The resonance magnetic field $B_0$ was determined from the sharp edge of the atomic loss feature~\cite{naka}.
The doublet structure arising from the resonance shift between different angular momentum projections $m_{l}=0$ and $m_{l}=\pm1$ appears in a $p$-wave Feshbach resonance due to dipole-dipole interactions between the two atoms in a $p$-wave molecular state~\cite{ticknor}. We treated the $p$-wave Feshbach resonance as a single resonance due to overlap of the $m_{l}=0$ and $m_{l}=\pm1$ resonances in $^6$Li~\cite{chevy}.
\begin{figure}[tbp]
\begin{tabular}{@{}cccc@{}}
\includegraphics[width=3.250 in]{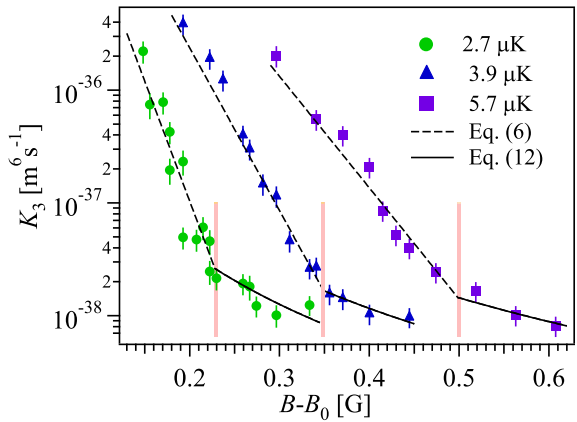}
\end{tabular}
\caption{Three-body loss coefficient $K_{3}$ versus magnetic field detuning $B-B_{0}$ at $T=2.7~\mu$K (solid circles), $T=3.9~\mu$K (solid triangles), and $T=5.7~\mu$K (solid squares). The vertical bars correspond to the point where $k_{\rm T}/k_{\rm r} \approx 0.3$. The black solid lines show the universal scaling of $K_{3} \propto V_{p}^{8/3}$ when $k_{\rm T}/k_{\rm r} < 0.3$. The dashed curves show the result of Eq.~(\ref{k3fit2}) when $k_{\rm T}/k_{\rm r} > 0.3$.} 
\label{fig:mag}
\end{figure}

Near the $p$-wave Feshbach resonance, the density of atom decay due to three-body recombination losses is governed by the following rate equation:
\begin{equation}
\frac{dn}{dt}= - K_{3} n^{3} -\Gamma n,
\label{raten}
\end{equation}
where, $\Gamma$ is the one-body loss rate.
Assuming that the atomic density profile is Gaussian, integrating Eq.~(\ref{raten}) results in a rate equation that describes the decay of the total number of atoms, $N$,
\begin{equation}
\frac{d N}{dt}= - K_{3} \left\langle n^2 \right\rangle N-\Gamma N,
\label{rate}
\end{equation}
where $\left\langle n^2 \right\rangle = (N^{2}/\sqrt{27}) \left[m \tilde{\omega}^{2} / (2 \pi k_{\rm B} T) \right]^{3}$ is the mean-square atomic density, which we determined from the atomic density profile obtained from the absorption image. Here, $\tilde{\omega}=(\omega_{x} \omega_{y} \omega_{z})^{1/3}$ is the mean trapping frequency.

Figure~\ref{fig:mag} shows the three-body loss coefficient $K_{3}$ as a function of the magnetic field detuning $B-B_{0}$ at $T=2.7~\mu$K and $n = 1.2 \times 10^{18} $m$^{-3}$ (solid circles), $T=3.9~\mu$K and $n = 1.3 \times 10^{18} $m$^{-3}$ (solid triangles), and $T=5.7~\mu$K and $n = 1.5 \times 10^{18} $m$^{-3}$ (solid squares), respectively. 
The vertical bars indicate the magnetic field detuning values at which $k_{\rm T}/k_{\rm r} \approx 0.3$. The dashed curves in Fig.~\ref{fig:mag} show the results of Eq.~(\ref{k3fit2}), which agree with the experimental data to the left of the vertical bars. To the right of the vertical bars, where $k_{\rm T}/k_{\rm r} \leq 0.3$ (or equivalently ($k_{\rm T}/k_{\rm r})^{2} \leq 0.1 $), the measured three-body loss coefficients show a different trend from Eq.~(\ref{k3fit2}). Therefore, the direct three-body recombination process dominates over the indirect three-body loss coefficient at small scattering volumes. 
In this small scattering volume limit, when $k_{\rm T}/k_{\rm r} \leq 0.3$~\cite{yoshida}, the three-body loss coefficient $K_{3}$ follows the universal scaling law of
\begin{equation}
K_{3}=C \frac{\hbar}{m} k_{\rm T}^{4} V_{p}^{8/3},
\label{law3d}
\end{equation}
as shown by the black solid curves with dimensionless constant $C=2 \times 10^{6}$ ~\cite{yoshida} on the right side of the vertical bars. The above scaling law was first predicted by Suno \textit{et al}.,~\cite{suno}, and was experimentally confirmed in our previous work~\cite{yoshida}.
During fitting, we fit the data of Fig.~\ref{fig:mag} to Eq.~(\ref{k3fit2}) in the nonuniversal regime (left side of the vertical bars) and to Eq.~(\ref{law3d}) in the universal scaling regime (right side of the vertical bars) with $K_{AD}$ and the critical magnetic field detuning $\delta B_{c}=B-B_{0}$ as free parameters at each temperature independently. 
We obtained consistent values of $K_{AD}$ from all data sets, which gave $K_{AD} \approx (1.3 \pm 0.5) \times 10^{-15}$~m$^{3}/$s from the average of all fits. The values of $\delta B_{c}$ at the kink location consistently correspond to $k_{\rm T}/k_{\rm r} \approx 0.30 \pm 0.01$. 
In our case, the ratio $(\Gamma_{\rm r}/ \hbar) / (n K_{AD}) \gtrsim 4$, which indicates that three-body losses are due to the two successive processes of pair creation and collision with a third atom above the universal scaling regime described by Eq.~(\ref{law3d}).

The quantity $k_{\rm T}/k_{\rm r}$ is the ratio of the second term to the first term in Eq.~(\ref{range3d}). 
Revamping Eq.~(\ref{range3d}) ~\cite{shotan} under the condition of $V_{p}^{-1} > k_{\rm e} k^{2}$, gives
\begin{equation}
-\frac{\tan \delta_{p}(k)}{k}= V_{p} k^{2} + k_{\rm e} V_{p}^{2} k^{4}.
\end{equation}
Multiplying the above expression with effective range $k_{\rm e}$, using the definition $k_{\rm r}=1/\sqrt{V_{p} k_{\rm e}}$, and considering $k \equiv k_{\rm T}$ (because generally the wave vector $k$ shows the temperature dependence~\cite{russian}), we obtained a convenient form of the effective range expansion:
\begin{equation}
\tilde{\delta_{p}}=-\frac{k_{\rm e} \tan \delta_{p}(k)}{k_{\rm T}} \approx \left(\frac{k_{\rm T}}{k_{\rm r}}\right)^{2}-\left(\frac{k_{\rm T}}{k_{\rm r}}\right)^{4}.
\end{equation}
Hence, $k_{\rm T}/k_{\rm r}$ includes the contribution of the effective range term in the collisional phase shift and quantifies the closeness of the dimer bound-state energy level to the average collision energy of two atoms.
In the limit of small $k_{\rm T}/k_{\rm r}$ in the far-resonant regime, the strength of the three-body losses due to Eq.~(\ref{k3fit2}) falls exponentially and becomes negligible compared to the direct three-body recombination losses described by Eq.~(\ref{law3d}). As a result, the three-body losses due to the universal scaling of Eq.~(\ref{law3d}) increase, starting to dominate at $k_{\rm T}/k_{\rm r} \approx 0.3$ in our case. Furthermore, the effect of the effective range $k_{\rm e}$ that is included in Eq.~(\ref{k3fit2}) becomes irrelevant when $k_{\rm T}/k_{\rm r} \leq 0.3$, which is consistent with previous observations~\cite{suno, yoshida}. Our current theoretical description of the three-body loss coefficient above the universal scaling regime converges to a single physical parameter $K_{AD}$ instead of two unknown parameters $\beta$ and $\gamma$~\cite{yoshida}.
\begin{figure}[tbp]
\begin{tabular}{@{}cccc@{}}
\includegraphics[width=3.250 in]{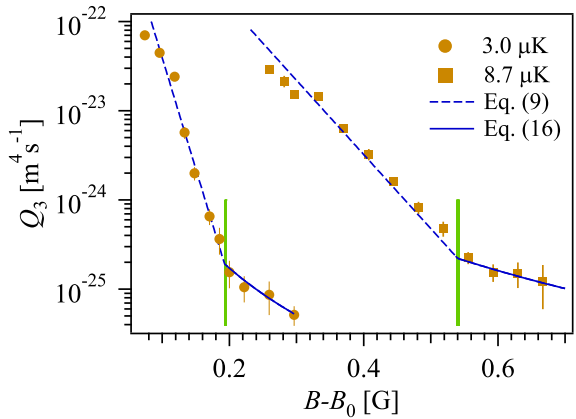}
\end{tabular} 
\caption{Three-body loss coefficient $Q_{3}$ versus magnetic field detuning $B-B_{0}$ in two dimensions at $T =3.0~\mu$K and $\omega_{z} = 2 \pi \times 168$~kHz (solid circles), and at $T =8.7~\mu$K and $\omega_{z} = 2 \pi \times 350$~kHz (solid squares).
The vertical bars correspond to values of $\delta B_{c}=B-B_{0}$ at which $q_{\rm T}/q_{\rm r} \approx 0.3$. The blue solid curves show a universal scaling of $Q_{3} \propto A_{p}^{3}$ when $q_{\rm T}/q_{\rm r} \leq 0.3$. The dashed curves show the result of Eq.~(\ref{Q3fit}) when $q_{\rm T}/q_{\rm r} > 0.3$.}
\label{fig:mag2d}
\end{figure}

\emph{Quasi-2D case:}
We prepared a 2D Fermi gas by adiabatically shining an optical lattice potential using a laser with a beam waist of 65~$\mu$m at 1064~nm, as described elsewhere~\cite{waseem,waseem2}. 
We calibrated the lattice depth from measured radial trap frequencies with and without the retroreflection beam. 
The motion of atoms along the lattice direction is restricted because the axial energy $\hbar \omega_z$ is always kept higher than the thermal energy $k_{\rm B} T$, where $\omega_z$ is the confinement frequency along the lattice direction.
To make sure that all the atoms are in the lowest motional state in the lattice potential, we performed the adiabatic band-mapping technique~\cite{waseem,kohl}.
We verified that the percentage of the population in the first excited motional state along the $z$ axis is less than 10$\%$ by taking into account both the size of the Brillouin zone and our imaging resolution~\cite{miranda}. 
In our experiment, the magnetic field is parallel to the optical lattice beams. Therefore, only the $p$-wave Feshbach resonance for $m_{l}=\pm 1$ symmetry becomes active because all atoms are in the motional ground state along the quantization direction~\cite{gun, waseem, tan}.

To extract the three-body loss constant $Q_{3}$ in the quasi-2D system, the time evolution of the number of atoms per site, $N_{s}$, is fitted to the solution of the rate equation, 
\begin{equation}
\frac{d N_{s}}{dt}= - Q_{3} \left\langle n_{s}^2 \right\rangle N_{s}-\Gamma N_{s},
\label{rate2d}
\end{equation}
where $\left\langle n_{s}^{2} \right\rangle=(N_{s}^{2}/3) \left[ m \tilde{\omega}^{2}/2 \pi  k_{\rm B} T\right]^2$ is the mean-square atomic density per site with 2D mean trap frequency $\tilde{\omega}=\sqrt{\omega_{x} \omega_{y}}$~\cite{miranda}. 
We performed all measurements in a thermal regime so that we can safely assume the atomic density profile as a Gaussian along the loosely confined axes.
Solid circles in Fig.~\ref{fig:mag2d} show $Q_{3}$ as a function of magnetic field detuning $B-B_{0}$ at $T = 3.0$~$\mu$K, $n_{s} = 7.0 \times 10^{11}~$m${^{-2}}$, and with a lattice depth of $V_{\rm lat} = 8.0$ $E_{\rm rec}$ ($\omega_{z} = 2 \pi \times 168$~kHz).
Here, $E_{\rm{rec}}=\hbar^2 k^2/(2m) \approx k_{\rm B} \times 1.4$~$\mu$K is the recoil energy of the $^{6}$Li atom.
Similarly, in Fig.~\ref{fig:mag2d}, solid squares indicate $Q_{3}$ values $T = 8.7$~$\mu$K, $n_{s} = 9.7 \times 10^{11}$m${^{-2}}$, and with a lattice depth of $V_{\rm lat} = 35.0$~$E_{\rm rec}$ ($\omega_{z} = 2 \pi \times 350$~kHz).
The vertical bars indicate the values of the magnetic field detuning $B-B_{0}$ at which $q_{\rm T}/q_{\rm r} \approx 0.3$.
In Fig.~\ref{fig:mag2d}, the dashed curves on the left side of the vertical bar show the theoretical results obtained from Eq.~(\ref{Q3fit}) with $Q_{AD} \approx (8.3 \pm 1) \times 10^{-9}$~m$^{2}/$s, in agreement with experimental data.
The ratios $(\gamma_{\rm r}/ \hbar) / ( n_{s} Q_{AD}) \approx 3$ for Fig.~\ref{fig:mag2d}(a) and $(\gamma_{\rm r}/ \hbar) / ( n_{s} Q_{AD}) \approx 6$ for Fig.~\ref{fig:mag2d}(b) again indicate that the observed three-body losses are mainly due to indirect two-step processes when $q_{\rm T}/q_{\rm r} \geq 0.3$.

In the region where $q_{\rm T}/q_{\rm r} \leq 0.3$, the experimental data deviate from the indirect process given by Eq.~(\ref{Q3fit}).
Therefore, we expect some universal scaling of the three-body loss rate coefficient $Q_{3}$ similar to the 3D case.
In the far-resonant region, the temperature threshold law can be written as $Q_{3} \propto T^{\lambda_{\rm min}}$~\cite{jpd2d}. Here, $\lambda_{\rm min}=2$ is a quantum number associated with the dominated lowest three-body continuum channel in the low-energy limit for three identical fermions in two dimensions~\cite{jpd2d}.
Therefore, using the dimensional analysis together with the temperature threshold law similar to the 3D case, we expect the following scaling law:
\begin{equation}
Q_{3}= \eta \frac{\hbar}{m} q_{\rm T}^{4} A_{p}^{3}.
\label{law2d}
\end{equation}
In analogy with $s$-wave interactions~\cite{kra}, the dimensionless constant $\eta$ may have additional scattering area dependence. The super Efimov effect~\cite{nishida} might be reflected in a universal double-exponential scaling behavior of $\eta$. 
We did not observe recombination loss peaks arising from triatomic super Efimov resonances as predicted for three identical fermions in the $m_{l}=\pm 1$ state confined in two dimensions~\cite{nishida}. Therefore, we consider $\eta$ a dimensionless constant as long as the threshold law and scattering area scaling law hold.
\begin{figure}[tbp]
\includegraphics[width=3.250 in]{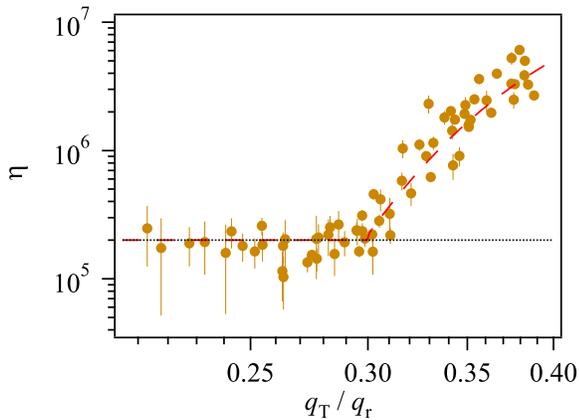} 
\caption{(a) Dimensionless parameter $\eta$ versus $q_{\rm T}/q_{\rm r}$. The dotted horizontal line shows a constant value of $\eta \approx 2  \times 10^{5}$, which demonstrates the expected scaling law of Eq.~(\ref{law2d}) below $q_{\rm T}/q_{\rm r}=0.3$.
The long dashed curve shows the results of Eq.~(\ref{Q3fit}), which describes the deviation of the data from the scaling of Eq.~(\ref{law2d}) at $q_{\rm T}/q_{\rm r} \geq 0.3$.}
\label{fig:law2d}
\end{figure}

In Fig.~\ref{fig:mag2d}, the blue solid curves show the scaling of $A_{p}^{3}$ at fixed temperature using $\eta \approx 2 \times 10^{5}$ in Eq.~(\ref{law2d}).
In our experiment, we were restricted to confirming the threshold law of $T^{2}$ in two dimensions at fixed $A_{p}$ because it was not possible to introduce controlled changes in temperature at a fixed lattice depth~\cite{comment}.
Therefore, we followed the method of displaying all results in dimensionless units~\cite{yoshida}. We plot $\eta$ as a function of $q_{\rm T}/q_{\rm r}$ in Fig.~\ref{fig:law2d} to clearly demonstrate the scaling of Eq.~(\ref{law2d}) and the range over which it applies. 
The entire data set along with additional data taken at 2.0, 3.4, 4.7, and $6.6~\mu$K merge into a single monotonic curve.  
Below $q_{\rm T}/q_{\rm r}=0.3$, the flat region of zero slope with an intercept of $2 \times 10^{5}$ best revealed the expected scaling law of Eq.~(\ref{law2d}) as shown by the dotted horizontal line in Fig.~\ref{fig:law2d}.
The long dashed curve in Fig.~\ref{fig:law2d} is the combined result from Eqs.~(\ref{law2d}) and~(\ref{Q3fit}).
The scaling of Eq.~(\ref{law2d}) holds over a relatively small range in the weakly interacting limit.
In our experiment, the lower limit of Eq.~(\ref{law2d}) scaling is restricted to $q_{\rm T}/q_{\rm r}=0.2$, mainly due to the non-negligible contribution of one-body decay.
After the upper limit at $q_{\rm T}/q_{\rm r}=0.3$, the loss coefficient steeply increased similar to the 3D case, and was well described by Eq.~(\ref{Q3fit}).

\section{Conclusion}
In conclusion, we have quantitatively analyzed the three-body loss coefficient as a function of temperature and magnetic field using a theoretical model based on a two-step loss process. The model nicely explains the loss behavior just above the weak interaction limit, where the universal scaling law is expected to hold.
The model reproduces the experimental three-body loss coefficient results in both 3D and quasi-2D geometries with a single parameter, i.e., an atom-molecule collision rate coefficient. 
In the weak-interaction limit, the three-body loss coefficient is explained by the scaling law, meaning that the three-body loss is dominated by the direct transition from the three free-atom-state into a deeply bound dimer and one free atom. Slightly above the scaling-law regime, the three-body loss coefficient deviates from the scaling law but the behavior can be explained by the model presented here, indicating that the universal scaling law breaks down because the Feshbach dimer creation gets faster and the three-body loss can be considered a two-step process of a Feshbach dimer creation and an atom-dimer collision. 

Our results provide useful initial grounds for extending the understanding of three-body physics in the presence of $p$-wave Feshbach resonances. In our current theoretical description, the atom-molecule collision rate coefficient is the only parameter and is theoretically more difficult to predict. We assumed that the atom-molecule rate coefficient is independent of the magnetic field due to the nonresonant nature of the relaxation process. However, its precise dependence on atom-molecule collision energy or temperature would be expected to vary over a wide range of temperatures, from a very few to a few hundred microkelvins. 
Understanding the atom-molecule rate coefficients as a function of collision energy near the $p$-wave Feshbach resonance in both 3D and 2D cases is needed to fully understand atomic losses near the Feshbach resonance, which prevents the realization of the $p$-wave superfluid.

\section*{ACKNOWLEDGMENT}
This work was supported by a Grant-in-Aid for Scientific Research on Innovative Areas (Grant No. 24105006) and a Grant-in-Aid for Challenging Exploratory Research (Grant No. 17K18752). M.W. acknowledges the support of a Japanese government scholarship (MEXT).


\begin{thebibliography}{90}
\bibliographystyle{apsrev4-1}

\bibitem{vg} V. Gurarie, L. Radzihovsky, and A. V. Andreev, \href{https://journals.aps.org/prl/abstract/10.1103/PhysRevLett.94.230403} {Phys. Rev. Lett. \textbf{94}, 230403 (2005)}.

\bibitem{cheng} C. -H. Cheng and S. -K. Yip, \href{https://journals.aps.org/prl/abstract/10.1103/PhysRevLett.95.070404} {Phys. Rev. Lett. \textbf{95}, 070404 (2005)}.

\bibitem{sukjin} S. Yoon and G. Watanabe \href{https://journals.aps.org/prl/abstract/10.1103/PhysRevLett.119.100401} {Phys. Rev. Lett. \textbf{119}, 100401 (2017)}.

\bibitem{victor} V. Gurarie and L. Radzihovsky, \href{https://www.sciencedirect.com/science/article/pii/S0003491606002399} {Ann. Phys. \textbf{322}, 2 (2007)}.

\bibitem{lev} J. Levinsen, N. R. Cooper, and V. Gurarie, \href{https://journals.aps.org/pra/abstract/10.1103/PhysRevA.78.063616} {Phys. Rev. A. \textbf{78}, 063616 (2008)}.

\bibitem{lev11} D. Vollhardt and P. Wolfe, \textit{The Superfluid Phases Of Helium 3} (Taylor and Francis Ltd, N.Y., 2002).


\bibitem{regal} C. A. Regal, C. Ticknor, J. L. Bohn, and D. S. Jin \href{https://journals.aps.org/prl/abstract/10.1103/PhysRevLett.90.053201}{Phys. Rev. Lett. \textbf{90}, 053201 (2003).}

\bibitem{zhang} J. Zhang, E. G. M. van Kempen, T. Bourdel, L. Khaykovich, J. Cubizolles, F. Chevy, M. Teichmann, L. Tarruell, S. J. J. M. F. Kokkelmans, and C. Salomon,\href{https://journals.aps.org/pra/abstract/10.1103/PhysRevA.70.030702} {Phys. Rev. A \textbf{70}, 030702 (2004)}.


\bibitem{ticknor} C. Ticknor, C. A. Regal, D. S. Jin, and J. L. Bohn, \href{https://journals.aps.org/pra/abstract/10.1103/PhysRevA.69.042712}{Phys. Rev. A. \textbf{69}, 042712 (2004)}.

\bibitem{chevy} F. Chevy, E. G. M. van Kempen, T. Bourdel, J. Zhang, L. Khaykovich, M. Teichmann, L. Tarruell, S. J. J. M. F. Kokkelmans, and C. Salomon, \href{https://journals.aps.org/pra/abstract/10.1103/PhysRevA.71.062710} {Phys. Rev. A. \textbf{71}, 062710 (2005)}.

\bibitem{gun} K. G\"{u}nter, T. St\"{o}ferle, H. Moritz, M. K\"{o}hl, and T. Esslinger, \href{https://journals.aps.org/prl/abstract/10.1103/PhysRevLett.95.230401}{Phys. Rev. Lett. \textbf{95}, 230401 (2005)}.

\bibitem{fed} A. K. Fedorov, V. I. Yudson, and G. V. Shlyapnikov, \href{https://journals.aps.org/pra/abstract/10.1103/PhysRevA.95.043615} {Phys. Rev. A \textbf{95}, 043615 (2017)}.

\bibitem{russian} D. V. Kurlov and G. V. Shlyapnikov, \href{https://journals.aps.org/pra/abstract/10.1103/PhysRevA.95.032710} {Phys. Rev. A \textbf{95}, 032710 (2017)}.

\bibitem{waseem2} M. Waseem, T. Saito, J. Yoshida, and T. Mukaiyama, \href{https://journals.aps.org/pra/abstract/10.1103/PhysRevA.96.062704} {Phys. Rev. A \textbf{96}, 062704 (2017).}

\bibitem{yoshida} J. Yoshida, T. Saito, M. Waseem, K. Hattor, and T. Mukaiyama; \href{https://journals.aps.org/prl/pdf/10.1103/PhysRevLett.120.133401} {Phys. Rev. Lett. \textbf{120}, 133401 (2018)}.

\bibitem{waseem3} M. Waseem, J. Yoshida, T. Saito, T. Mukaiyama \href{https://journals.aps.org/pra/abstract/10.1103/PhysRevA.98.020702}{Phys. Rev. A \textbf{98}, 020702(R), (2018)}.

\bibitem{elk} A. Crubellier, R. G-Férez, C. P. Koch, and E. L-Koenig \href{https://journals.aps.org/pra/abstract/10.1103/PhysRevA.99.032710}{Phys. Rev. A \textbf{99}, 032710, 2019}


\bibitem{gab} J. P. Gaebler, J. T. Stewart, J. L. Bohn, and D. S. Jin, \href{https://journals.aps.org/prl/abstract/10.1103/PhysRevLett.98.200403}{Phys. Rev. Lett. \textbf{98}, 200403 (2007)}.

\bibitem{inada} Y. Inada, M. Horikoshi, S. Nakajima, M. Kuwata-Gonokami, M. Ueda and T. Mukaiyama, \href{https://journals.aps.org/prl/abstract/10.1103/PhysRevLett.101.100401}{Phys. Rev. Lett. \textbf{101}, 100401 (2008)}.

\bibitem{waseem} M. Waseem, Z. Zhang, J. Yoshida, K. Hattori, T. Saito, and T. Mukaiyama, \href{http://iopscience.iop.org/article/10.1088/0953-4075/49/20/204001/meta} {J. Phys. B \textbf{49}, 204001 (2016)}.

\bibitem{fuch} J. Fuchs, C. Ticknor, P. Dyke, G. Veeravalli, E. Kuhnle, W. Rowlands, P. Hannaford, and C. J. Vale, \href{https://journals.aps.org/pra/abstract/10.1103/PhysRevA.77.053616} {Phys. Rev. A. \textbf{77}, 053616 (2008)}.

\bibitem{naka} T. Nakasuji, J. Yoshida and T. Mukaiyama, \href{https://journals.aps.org/pra/abstract/10.1103/PhysRevA.88.012710} {Phys. Rev. A. \textbf{88}, 012710 (2013)}.


\bibitem{ueda1} S. M. Yoshida and M. Ueda, \href{https://journals.aps.org/prl/abstract/10.1103/PhysRevLett.115.135303}{Phys. Rev. Lett. \textbf{115}, 135303 (2015)}.

\bibitem{thy} C. Luciuk, S. Trotzky, S. Smale, Z. Yu, S. Zhang, and J. H. Thywissen, \href{https://www.nature.com/articles/nphys3670} {Nature. Phys \textbf{12}, 599 (2016)}.


\bibitem{yao} J. Yao and S. Zhang,\href{https://journals.aps.org/pra/abstract/10.1103/PhysRevA.97.043612}{Phys. Rev. A \textbf{97}, 043612 (2018)}.

\bibitem{lio} G. Liu and Y.-C. Zhang,\href{https://iopscience.iop.org/article/10.1209/0295-5075/122/40006/meta}{ Europhys. Lett. \textbf{122}, 40006
(2018).}



\bibitem{esry1} B. D. Esry, Chris H. Greene, and H. Suno, \href{https://journals.aps.org/pra/abstract/10.1103/PhysRevA.65.010705} {Phys. Rev. A \textbf{65}, 010705(R) (2001)}.


\bibitem{suno} H. Suno, B. D. Esry, and C. H. Greene, Phys. \href{https://journals.aps.org/prl/abstract/10.1103/PhysRevLett.90.053202} {Phys. Rev. Lett. \textbf{90}, 053202 (2003)}.

\bibitem{suno2} H. Suno, B. D. Esry, and C. H. Greene, \href{http://iopscience.iop.org/article/10.1088/1367-2630/5/1/353} {New J. Phys. \textbf{5}, 53 (2003)}.

\bibitem{jona} M. Jona-Lasinio, L. Pricoupenko, and Y. Castin, \href{https://journals.aps.org/pra/abstract/10.1103/PhysRevA.77.043611} {Phys. Rev. A \textbf{77}, 043611 (2008)}.


\bibitem{jesper} J. Levinsen, N. R. Cooper, V. Gurarie, \href{https://journals.aps.org/prl/abstract/10.1103/PhysRevLett.99.210402} {Phys. Rev. Lett. \textbf{99}, 210402 (2007)}.

\bibitem{jpd} J. P. D'Incao, B. D. Esry, and C. H. Greene, \href{https://journals.aps.org/pra/abstract/10.1103/PhysRevA.77.052709} {Phys. Rev. A \textbf{77}, 052709 (2008)}.

\bibitem{bratpra} E. Braaten, P. Hagen, H.-W. Hammer, and L. Platter, \href{https://journals.aps.org/pra/abstract/10.1103/PhysRevA.86.012711}{Phys. Rev. A \textbf{86}, 012711 (2012)}.

\bibitem{nishida} Y. Nishida, S. Moroz, and D. T. Son,  \href{https://journals.aps.org/prl/abstract/10.1103/PhysRevLett.110.235301}{Phys. Rev. Lett. \textbf{110}, 235301 (2013)}.

\bibitem{nishida12} Y. Nishida and S. Tan, Few-Body Syst. \textbf{51}, 191 (2011).
\bibitem{nishida13} T. K. Lim and P. A. Maurone, \href{https://journals.aps.org/prb/abstract/10.1103/PhysRevB.22.1467}{Phys. Rev. B \textbf{22}, 1467 (1980)}.
\bibitem{nishida14} Y. Nishida, \href{https://journals.aps.org/pra/abstract/10.1103/PhysRevA.86.012710}{Phys. Rev. A \textbf{86}, 012710 (2012).}

\bibitem{nikolaj}  A. G. Volosniev, D. V. Fedorov, A. S. Jensen, N. T. Zinner, \href{http://iopscience.iop.org/article/10.1088/0953-4075/47/18/185302/meta}{J. Phys. B \textbf{47}, 185302 (2014).}

\bibitem{marish} V. Ngampruetikorn, M. M. Parish, and J. Levinsen, \href{http://iopscience.iop.org/article/10.1209/0295-5075/102/13001/meta}{Europhys. Lett. \textbf{102}, 13001 (2013)}.

\bibitem{jpd2d} J. P. D'Incao and B. D. Esry, \href{https://journals.aps.org/pra/abstract/10.1103/PhysRevA.90.042707} {Phys. Rev. A \textbf{90}, 042707 (2014)}.

\bibitem{zen1} C. Gao, J. Wang, and Z. Yu, \href{https://journals.aps.org/pra/abstract/10.1103/PhysRevA.92.020504}{Phys. Rev. A \textbf{92}, 020504 (2015)}.

\bibitem{zen2} P. Zhang and Z. Yu, \href{https://journals.aps.org/pra/abstract/10.1103/PhysRevA.95.033611}{Phys. Rev. A \textbf{95}, 033611 (2017)}.

\bibitem{maxim} M. A. Efremov, L. Plimak, M. Y. Ivanov, and W. P. Schleich, \href{https://journals.aps.org/prl/abstract/10.1103/PhysRevLett.111.113201}{Phys. Rev. Lett. \textbf{111}, 113201 (2013)}.


\bibitem{sjj} S.-J. Jiang and F. Zhou, \href{https://journals.aps.org/pra/abstract/10.1103/PhysRevA.97.063606}{Phys. Rev. A \textbf{97}, 063606 (2018)}.

\bibitem{hui}  H. Hu, B. C. Mulkerin, L. He, J. Wang, X. -J. Liu, \href{https://journals.aps.org/pra/abstract/10.1103/PhysRevA.98.063605}{Phys. Rev. A \textbf{98}, 063605 (2018)}

\bibitem{petrov2}  B. Bazak, D. S. Petrov, \href{https://journals.aps.org/prl/abstract/10.1103/PhysRevLett.121.263001}{Phys. Rev. Lett. \textbf{121}, 263001 (2018)}.







\bibitem{Taylor} J. R. Taylor, \textit{Scattering Theory} (Wiley, New York, 1972).
\bibitem{naidon}  P. Zhang, P. Naidon, and M. Ueda, \href{https://journals.aps.org/pra/abstract/10.1103/PhysRevA.82.062712}{Phys. Rev. A \textbf{82}, 062712 (2010)}.


\bibitem{idz} Z. Idziaszek, \href{https://journals.aps.org/pra/abstract/10.1103/PhysRevA.79.062701}{Phys. Rev. A 79, 062701 (2009)}.

\bibitem{lee} J. Li, J. Liu, L. Luo, and B. Gao \href{https://journals.aps.org/prl/abstract/10.1103/PhysRevLett.120.193402}{Phys. Rev. Lett. \textbf{120}, 193402 (2018)}



\bibitem{cai} Y.-C. Zhang and S. Zhang, \href{https://journals.aps.org/pra/abstract/10.1103/PhysRevA.95.023603} {Phys. Rev. A \textbf{95}, 023603 (2017)}.

\bibitem{prico} L. Pricoupenko, \href{https://journals.aps.org/prl/abstract/10.1103/PhysRevLett.100.170404}  {Phys. Rev. Lett. \textbf{100}, 170404 (2008)}.


\bibitem{shotan} Z. Shotan, O. Machtey, S. Kokkelmans, and L. Khaykovich, \href{https://journals.aps.org/prl/abstract/10.1103/PhysRevLett.113.053202}{Phys. Rev. Lett. \textbf{113}, 053202 (2014)}.


\bibitem{kohl} M. K\"{o}hl, H. Moritz, T. St\"{o}ferle, K. G\"{u}nter, and T. Esslinger, \href{https://journals.aps.org/prl/abstract/10.1103/PhysRevLett.94.080403}{Phys. Rev. Lett. \textbf{94}, 080403 (2005).}

\bibitem{miranda} M. H. G. de Miranda, A. Chotia, B. Neyenhuis, D. Wang, G. Quéméner, S. Ospelkaus, J. L. Bohn, J. Ye, and D. S. Jin, \href{https://www.nature.com/articles/nphys1939}{Nature Phys. \textbf{7}, 502 (2011)}.


\bibitem{tan} S. -G. Peng, S. Tan, and K. Jiang, \href{https://journals.aps.org/prl/abstract/10.1103/PhysRevLett.112.250401} {Phys. Rev. Lett. \textbf{112}, 250401 (2014)}.

\bibitem{kra} T. Kraemer , M. Mark , P. Waldburger , J. G. Danzl , C. Chin , B. Engeser , A. D. Lange , K. Pilch , A. Jaakkola, H.-C. N\"{a}gerl, R. Grimm, \href{https://www.nature.com/articles/nature04626} {Nature \textbf{440}, 04626 (2006)}.

\bibitem{comment} Inelastic collisions are sensitive to the confinement strength $l_{z}$ in the far resonance~\cite{russian}. Changing temperature in an optical dipole trap before loading into the optical lattice does not help in our experiment.















\end{thebibliography}
\end{document}